# Photonic Devices Based On Black-Phosphorus and Combined Hybrid 2D nano-materials

Leonardo Viti and Miriam S. Vitiello[*]

*NEST, Istituto Nanoscienze – CNR and Scuola Normale Superiore, Piazza San Silvestro 12, Pisa, I-56127*

[*]*miriam.vitiello@sns.it*

**Abstract**

Artificial semiconductor heterostructures played a pivotal role in modern electronic and photonic technologies, providing a highly effective mean for the manipulation and control of carriers, from the visible to the far-infrared. Despite the exceptional versatility, they commonly require challenging epitaxial growth procedures due to the need of clean and abrupt interfaces, which proved to be a major obstacle for the realization of room-temperature (RT), high-efficiency devices, like source, detectors or modulators. The discovery of graphene and the related fascinating capabilities have triggered an unprecedented interest in devices based on inorganic two-dimensional (2D) materials. Amongst them black-phosphorus (BP) recently showed an extraordinary potential in a variety of applications across micro-electronics and photonics. With an energy gap in-between the gapless graphene and the larger gap transition metal dichalcogenides, BP can form the basis for a new generation of high-performance photonic devices that could be engineered from "scratch" like transparent saturable absorbers, fast photocounductive switch and low noise photodetectors, exploiting its peculiar electrical, thermal and optical anisotropy. This paper will review the latest achievements in black phosphorus-based THz photonics and discuss future perspectives of this rapidly developing research field.

## 1. Introduction

Two-dimensional (2D) van der Waals layered materials [1, 2] display an extraordinary technological potential for engineering electronic and photonic devices and components; they also provide an intriguing platform for fundamental investigations, through the exploitation of their confined electronic systems. If placed on chip with integrated optical circuits [3], they can allow maximal interaction with light, therefore optimally utilizing their novel and versatile properties for a large number of applications in optical communications [4], spintronics [5], high-resolution gas sensing [6], tomography, ultrafast physics [7].

In the last few decade, graphene-oriented research [8] has had a dramatic research impact [9]. The superior carrier mobility, induced by the massless Dirac fermions, combines in graphene with a gapless spectrum that, although beneficial for applications requiring frequency-independent absorption [10, 11], also prevents the effective switching of its conductivity in electronic devices. This distinctive characteristic, allows charge carrier generation by light absorption over a very wide energy spectrum, while always inducing a significant amount of electricity. The inherent related "leakage" can significantly affect the efficiency of graphene-based devices. Finite and direct bandgaps materials are conversely desirable for a wealth of applications, including transparent optoelectronics, photovoltaics and photodetection. These issues are driving present research in the quest for alternative 2D layered materials, which, behaving like semiconductors, can only conduct electricity whenever the electrons absorb enough energy through heat, light, and other means. Depending on their specific band structures, these materials can disclose peculiar functionalities to be exploited for highly efficient light detection.

Single-unit-cell thick layers of transition-metal dichalcogenides (TMDCs: $MoS_2$, $MoSe_2$, $WS_2$, $WSe_2$, etc.) have recently emerged as a valuable solution [12]. 2D TMDCs can be obtained from bulk crystals by employing the micromechanical exfoliation method [9], like for the case of graphene, but they show a direct bandgap, ranging between ~0.4 eV and ~2.3 eV, which enables applications that well complement graphene capabilities [13]. In particular, 2D TMDCs are suitable for photovoltaic applications [2] and for devising robust ultra-thin-body field effect transistor (FET) architectures which can easily provide subthreshold swing of ~ 60 mV/dec and $I_{on}/I_{off}$ ratio up to $10^8$ [14]. Nonetheless, their relatively low mobility ($\leq$ 200 $cm^2V^{-1}s^{-1}$) is a major constraint for high-frequency electronic and photonic applications.

With an energy gap in-between the gapless graphene and the larger gap TMDCs, black phosphorus (BP) recently emerged as a fascinating and versatile material for high-frequency electronic and photonics applications [3, 15]. BP is the most thermodynamically stable allotrope of the phosphorus element. In standard conditions, it shows a layered graphite-like structure, where atomic planes are held together by weak Van der Waals forces of attraction, thus allowing the application of standard micromechanical exfoliation techniques. As in graphene, each atom in BP is connected to three neighbours, forming a stable layered honeycomb structure with an interlayer spacing of ~5.3 Å.

Unlike layered crystals with flat in-plane lattice, the hexagonally distributed phosphorus atoms are arranged in a puckered structure rather than in a planar one (Fig.1).

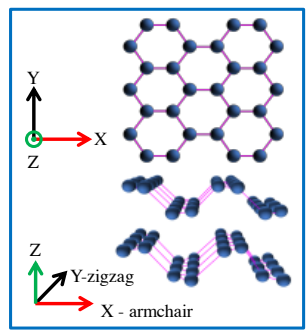

**Fig.1. Black phosphorus atomic structure**. BP atoms are arranged in puckered honeycomb layers bounded together by Van der Waals forces; the armchair (x) and zigzag (y) crystal axis are shown on the graph.

The BP monolayer is indeed puckered along the armchair (*x*) direction, therefore creating a fully anisotropic band structure that reflects in large electrical, thermal and visible/near IR-optical in-plane anisotropy [13, 16, 17, 18]. As a result, being an intrinsically flexible material, that converts heat in energy with high efficiency and with performances that does not require any sophisticated engineering, BP can effectively provide new functionalities in a variety of device applications.

Bulk BP has a small direct bandgap of ~0.3 eV, which enables the complete switching between insulating and conducting states when implemented in a transistor. The reduction of the flake thickness leads to quantum confinement which enhances the energy gap up to $E_g$ ~1.0 eV in the limit case of phosphorene (a single layer of BP); as a result the on-off current ratio ($I_{on}/I_{off}$) of a BP-based FET can be improved by employing thinner flakes: an $I_{on}/I_{off}$ ratio of ~$10^5$ has been recently reported in back-gated FET structures [13]. On the other hand, thickness reduction is detrimental for carrier mobility: thinner flakes are more vulnerable to scattering by interface impurities [19] and the effective mass of charge carriers increases when the number of atomic layers is reduced [20]. Despite this, BP thin films are endowed with hole mobilities exceeding 650 $cm^2V^{-1}s^{-1}$ at room temperature (RT) and well above 1000 $cm^2V^{-1}s^{-1}$ at 120 K [16], thus overtaking the limiting factor of large-gap TMDCs. For these reasons, BP represents an ideal material for applications in near and mid-infrared optoelectronics, high-speed thin film electronics, photonics and far-infrared optoelectronics. Indeed, it has successfully been employed in fast near-infrared photodetectors [3, 21, 22] and in radio-frequency FETs with oscillation frequencies up to 20 GHz [15]. Furthermore, the above-mentioned superb $I_{on}/I_{off}$ ratio of BP-based FETs makes BP well suited for detection of THz frequency light, being high-frequency operation and huge carrier density tunability undisputed benefits in the viewpoint of higher modulation frequency and larger sensitivities.

Exploiting black phosphorus and new hybrid materials combinations can open the path to ground-breaking implementations of active photonic devices and passive components across underexploited frequency ranges.

Here we will review the main achievements in this very recent but rapidly progressing field, and will discuss the opportunities for new performance improvements and further exciting discoveries.

## 2. Saturable Absorbers in the Telecom band – Ultrafast Nonlinear Photonics

Saturable absorbers (SA) are passive optical elements exhibiting an intensity-dependent transmission. Operating in transmission or in reflection (saturable absorber mirrors SESAMs), they are routinely used for driving solid-state lasers in the ultrafast mode-locked regime, turning the laser continuous-wave output into a train of short pulses, through the exploiting of a decreasing light absorption at increasing light intensities (absorption bleaching). Typical implementations of SAs rely on semiconducting materials, which under intense illumination are subjected to the saturation of their absorption, as a direct effect of valence band depletion, conduction band filling and ultra-fast (≈ 100 *fs*) intra-band or interband (*ps*) carrier thermalization [23] (Fig. 2). This effect can be qualitatively described as follows: when an electron in the valence band absorbs a photon having energy larger than the band-gap, it undergoes a transition to the conduction band and subsequently returns to the valence band by thermalization. However, if the number of photons is sufficiently large (high radiation intensity), the final electronic states in the valence and conduction bands will be fully occupied for a fraction of time shorter than the recombination lifetime, thus preventing further absorption (*Pauli-blocking*). Theoretically, this nonlinear effect is related to a third-order material nonlinearity arising from the imaginary part of the complex third-order susceptibility.

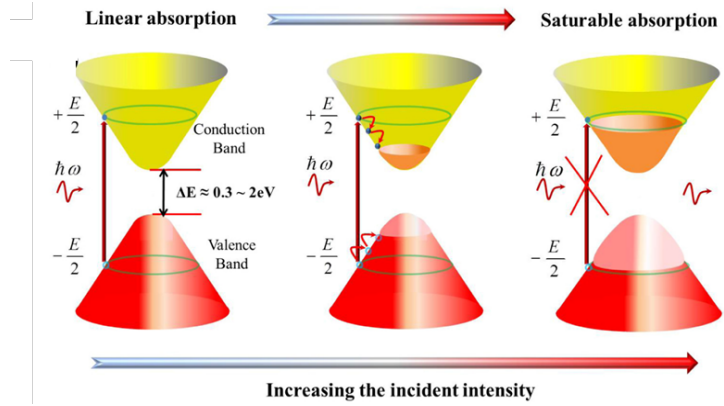

**Fig. 2. Saturable absorption**. Schematic diagram of the saturable absorption process.

Semiconductor-based SAs are typically wavelength selective and operate over a narrow band of the energy spectrum above their band-gap [23]. However, in practical applications, broadband (and tuneable) SAs are highly desirable. Furthermore, since the operation of SESAMs is typically restricted to wavelengths below 1500 nm, the search for new potential materials has recently become more eminent. In particular, 2D and 1D nanomaterials like graphene [24], single-walled carbon nanotubes [25], topological insulators [26] and TMDs [27, 28] have been widely investigated in the last years.

BP recently emerged as an ideal candidate, in this respect. Its tunable and thickness-dependent energy gap, in the wavelength range < 4.1 μm (corresponding to 0.3 eV), ensured by its orthorhombic crystal structure, opens unusual possibility for optical absorption over a broadband and tunable range of energies. If indeed $MoS_2$, a well-known TMD, also shows a bandgap that changes by varying its thickness, in contrast to any other material, BP always maintains a direct gap, regardless its thickness. This feature is of crucial importance to ensure efficient photoelectrical conversion and ultrafast photo-carrier dynamics, two fundamental benefits for infrared ultrafast photonics and high frequency optoelectronics. Moreover, the strong resonant absorption (> 5 %) and the low energy band gap can play a significant role for optical communications in the telecom band: BP shows a broad peak in the range between 1400 nm and 1600 nm, as can be seen from the absorption spectrum of solution-dispersed flakes (Fig.3).

When embedded in a laser cavity, SAs can induce passive mode-locking. In this latter case, SA dynamics occurring on ultrafast timescales, i.e. relaxation times shorter than the pulse duration, are usually preferred to maintain stable passive mode-locking in a laser. Very recent experimental investigations confirmed that this condition is well fulfilled in a few-layer BP, where the direct band gap ensures ultra-fast carrier relaxation dynamics (photocarrier lifetime ≈ 100 ps [29]). In the single-photon interaction picture, the absorption coefficient can then be described as consisting of two contributions:

$$\alpha(I) = \alpha_{ns} + \frac{\alpha_s}{1+ I/I_{sat}} \qquad (1)$$

where $\alpha_{ns}$ is the linear non-saturated absorption coefficient and $\alpha_s$ is the non-linear modulation loss of the SA, also termed modulation depth. The parameter $I_{sat}$ is defined as the optical intensity value at the point where the absorption coefficient is reduced to half of its original value. $\alpha_s$ and $I_{sat}$ are the two main figures of merits, defining the performance of the SA and are determined experimentally via the fit between the measured transmission $T_r$ through the absorber as a function of the radiation intensity and the relation:

$$Tr(I) = A \exp[-\alpha(I)] \qquad (2)$$

where A is a normalization coefficient. BP-based SAs have been recently demonstrated, showing broad frequency operation from the visible (400 nm – 640 nm) [30, 31] to the mid-infrared (1100 nm – 2100 nm) [31-34]. The SA characterization is conventionally performed via z-scan measurements: the absorber is placed on the light path of a focused beam and its position is varied along the propagation direction while the transmitted power is recorded. In this way, the radiation intensity impinging on the SA varies as the beam waist changes, reaching the maximum value at the focal point. The z-scan transmission experiments performed on a solution of few-layer BP and N-Methil pyrrolidone (NMP) at 1550 nm is shown in Fig.2b [32]. Normalized transmittance experiments as a function of the radiation intensity allows to retrieve the main figures of merits of the near IR SA, i.e. $\alpha_s$ = 9 % and $I_{sat}$ = 25 MW/cm$^2$ [32]

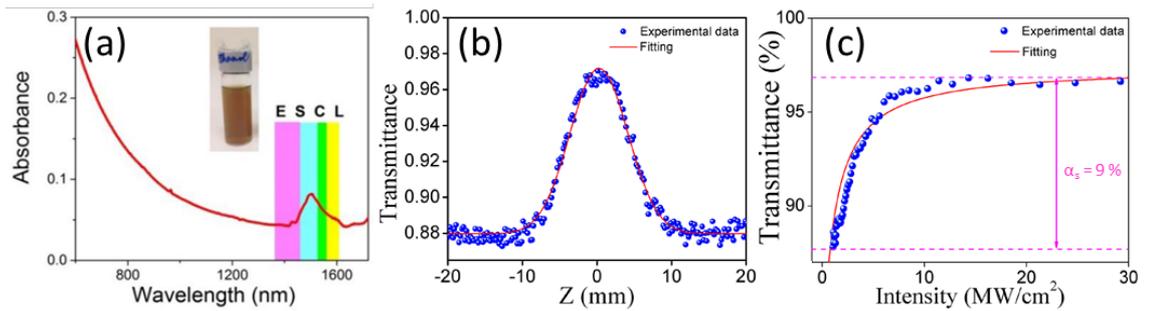

**Fig. 3. Saturable absorption in Black Phosphorus**. (a) UV-visible-infrared absorption spectrum of BP solution. The color bars indicate the telecommunication bands, i.e., E: extended communication band; S: short wavelength communication band; C: conventional communication band; L: long wavelength communication band. The inset shows the photograph of an undiluted BP dispersion in ethanol [34]. (b) z-scan curve of black phosphorus nano-sheets measured at 1550 nm and (c) the corresponding nonlinear saturable absorption curve [32].

The broadband tuneability of a few-layer BP makes such material an ideal candidate, also in the telecom range, very differently from TMDs, which have narrow absorption peaks in the telecom band caused by their edge states [35]. Moreover, the strong resonant absorption leads to high values of modulation depth (> 35 % [31]) much better than the graphene-based SAs, which are conversely limited by the relatively weak absorption of the material (2.3 % in the monolayer case). It has to be noted, however, that graphene typically presents much lower saturation intensities (≈ 2 MW/cm$^{-2}$), making it more sensitive to low irradiations then better suited for mode-locking applications.

The described attractive property of BP have been recently exploited to induce mode-locking in laser systems operating from the visible to the mid-IR [30-34], allowing to reach few-hundred femtosecond long pulses in the telecom range in a Erbium doped fiber (EDF) laser [33] (Fig.4a).
To really exploit the technological potential of BP saturable absorber to mode lock a laser, the material needs to be properly integrated along the beam path to reach a sufficiently strong light-matter interaction. The first and simplest configuration was realized with BP nanosheets obtained by liquid exfoliation in a solution of NMP deposited around the microfiber that acts as a waveguide for an EDF laser (Fig.4b). This scheme exploits evanescent field interaction of the propagating light with a few-layer BP. A modulation depth of 6.91 % was reported, providing laser pulse duration of 940 fs at a central wavelength of 1566.5 nm [32].

In order to improve the overlap between the electromagnetic field and the SA, more sophisticated strategies have been adopted to integrate the SA directly within the fiber. Sotor et *al.* [33] devised a laser system in which the SA is realized via mechanical exfoliation of ≈ 300 nm thick flakes on the fiber ferrule, in order to completely cover the fiber core. In this case, a maximum modulation depth of 4.6% was achieved, allowing to reach pulse widths of 272 fs. Alternatively, BP flakes [31] were exfoliated onto a quartz wafer then inserted in the resonant cavity of different solid state lasers operating at 639 nm, 1060 nm and 2100 nm.

Despite the demonstrated performance of BP saturable absorbers, the actual technological stability and feasibility is hindered by the material degradation under ambient exposure. To undertake this issue avoiding oxidation, Mu et *al.* [34] encapsulated multilayer BP nanoflakes within a polymer matrix, transparent in the telecom spectral range, to form a composite. The SA was made of robust membranes of BP-PVP (Poly-Vinyl-Pyrrolidone) nanofibers, prepared by using a semi-industrial electrospinning technique (Fig. 4c). With this active element Q-switched pulse generation has been observed at pump power of ~ 25 mW, i.e. at much lower thresholds than those of graphene [36] and quite similar to those obtained with TMDs [28] SAs. The performance of BP-PVP composite was found to be stable over several hours of operation, demonstrating the robustness of the devised approach.

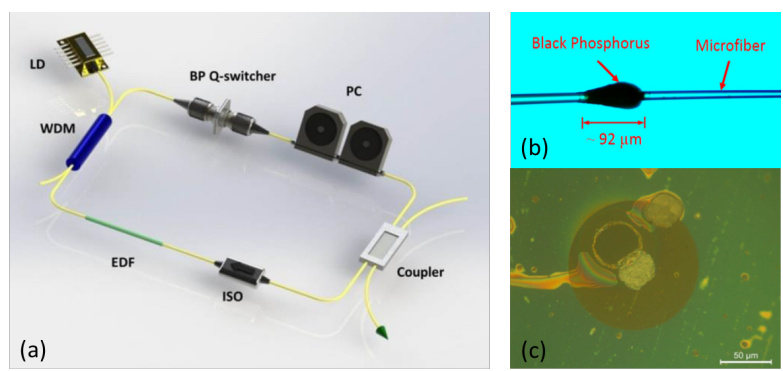

**Fig. 4. Q-switched fiber laser**. (a) Schematic illustration of the ring cavity of the Q-switched fiber laser realized in ref. [34]. Here LD is the 974 nm laser diode, ISO is a 1550 nm Polarization independent isolator, WDM is a wavelength division multiplexer, PC is a polarization controller and EDF is the erbium-doped fiber. (b) Optical image of a microfiber-based BP SA [32]. (c) Optical image of a PMMA-BP-PMMA composite film placed on the end facet of the optical fiber ferrule [34].

## 3. Photoconductive Switch in the visible telecom range

Photoconductive (PC) switches, also known as Auston switches, are semiconductor devices largely exploited for microwave [37] and terahertz optoelectronic applications [38]. Their working principle relies on the photogeneration of carriers via the absorption of photons having energy larger than the semiconductor band-gap. The sudden change in charge density results in a superb increase of the conductivity, which induces the device to operate like a switch. This effect, called *optical gating*, can be exploited to generate or receive a signal in the few GHz -THz range.

Conventionally these switches consist of transmission line center-fed antennas with a central gap bridged by the photoconductive material. The gap is then illuminated with a sub-picosecond laser pulse that *activates* the switch (Fig.5). Emitting photoconductive switches require a *dc* bias applied across the antenna. The electric signal generated by the excited photocurrent is radiated by the antenna with properties which depend from the semiconductor material and the antenna design. Carriers in the PC material react to the "primary" optical pulse within a certain time interval, given by the carrier recombination and momentum relaxation times. Then, the "secondary" pulse radiated by the antenna contains spectral signatures of these dynamics. In particular, carrier lifetimes in the order of few picoseconds allow for the generation of high frequency radiation, up to the terahertz frequency range. When operated as a receiver, the device is usually unbiased. Photocarriers, excited in the conduction band by the primary pulse, are accelerated by the incoming electromagnetic field that is funneled by the antenna on the photoconductive element. The induced current is then amplified and recorded.

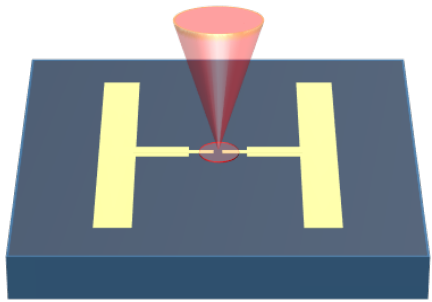

**Fig. 5. Schematic illustration of the microwave/THz switch**. The conductivity of the gap area between the two arms of the dipole antenna is activated upon illumination.

Fast and sensitive PCs require materials which once integrated in the antenna gap provide low dark currents and fast turn-off. The first condition is typically fulfilled by intrinsic, highly resistive materials, preferably having large energy gaps needed to avoid tunneling. The achievement of the latter condition is usually more demanding. The turn-off of a PC switch is the process through which the carrier population in conduction band drops off to the original (pre-illumination) concentration. The main process of carrier reduction is the electron-hole *recombination* and its duration is the material-dependent recombination lifetime ($\tau_{rec}$). A second process called *transit removal*, or *sweep-out*, accounts for the fact that carriers can reach the electrodes within a transit time ($\tau_{tr}$) that is equal to the distance between the contacts divided by the drift velocity. The efficiency of this process depends on the material mobility and contact design. In widely used semiconductor materials like silicon or GaAs, $\tau_{tr} < \tau_{rec}$, (in GaAs $\tau_{tr} \sim$ 1 ps, $\tau_{rec} \sim$ 100 ps [39]) thus the sweep-out process is faster than recombination. For high frequency switching applications, a material with high mobility (low $\tau_{tr}$) is then desirable.

If traditionally high mobility materials have been employed for high speed operation, low-mobility, low-temperature (LT) grown GaAs more recently replaced high-mobility GaAs thanks to its shorter carrier recombination lifetime, which allows for higher operation bandwidth and higher sensitivity. However, it can operate at wavelengths around 800 nm because the direct energy bandgap at 1.43 eV hampers its use for common optical communications around 1500 nm. Possible alternative strategies include III-V compounds and more complex materials like GaBiAs [40], GaNAsSb [41] and ion-implanted InGaAs [42, 43], having energy gaps < 0.75 eV and long wavelength operation up to 1600 nm. These devices represent the state of the art for ultrafast generation and detection in the THz range once using excitation lasers in the telecom band.

An alternative intriguing solution to extend the spectral bandwidth is represented by low band-gap 2D materials, like graphene and BP. The graphene gapless nature and the ultrafast carrier-carrier dynamics (10 fs), demonstrated to be ideal to devise graphene THz PC switches showing performance already comparable with commercial LT-GaAs devices [44]. On the other hand, although BP shows a photocarrier lifetime around 100 ps [29], detrimental for high-frequency operation, the low, tuneable and direct energy gap can play an important role for the realization of sensitive PC devices. The strong resonant absorption of BP, can indeed lead to small photon penetration depths (~ 300 nm for visible light [45]) and, consequently, to large carrier density modulations.

Despite the mentioned potential, only one report of BP PCs presently exists [46]. Microwave switches based on exfoliated BP and strongly responding to a 1.55 μm optical excitation have been indeed devised and realized in a coplanar waveguide geometry, where the gap between two linearly tapered contacts is bridged with a few-layers BP flake (Fig.6a).
The PC microwave performance is typically evaluated in terms of the ratio ($R_{ON/OFF}$) between the transmission through the switch with and without optical excitation. In the transfer matrix formalism the transmission coefficient through a two-terminal device is defined in terms of the scattering parameter $S_{21}$. Thus $R_{ON/OFF}$ is estimated from the following expression:

$$R_{ON/OFF} = \frac{S_{21}(ON)}{S_{21}(OFF)} \tag{3}$$

Fig.6b reports the ON/OFF ratio of the BP microwave switch as a function of the input frequency, for an optical excitation having 1.55 μm wavelength and 50 mW power. The four sets of data, collected for BP thickness ranging from 20 nm to "0" nm (bare silicon substrate) demonstrate the switching behavior of BP, with $R_{ON/OFF}$ = 17.2 dB at 1 GHz and $R_{ON/OFF}$ 5.7 dB at 10 GHz. If compared to the performances reported for III-V semiconductors ($R_{ON/OFF}$ = 4.5 dB at 10 GHz in GaNAsSb [41]), these results look promising for future implementation of BP in high sensitivity microwave switches operating in the telecom band.

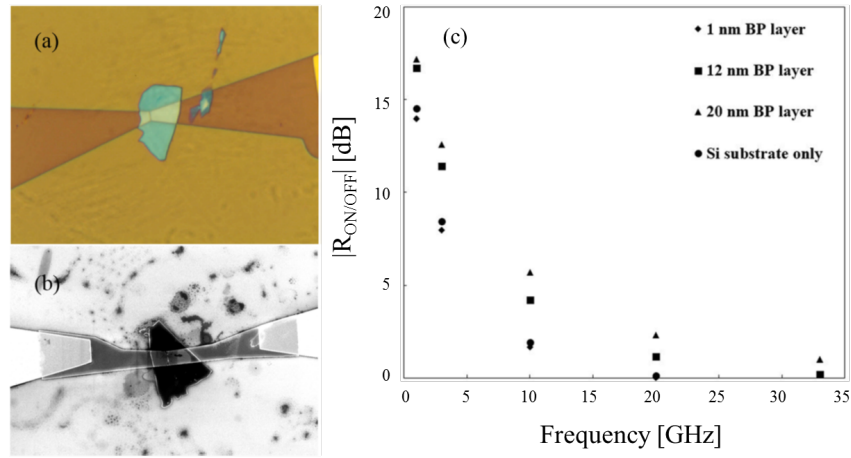

**Fig. 6. BP-based microwave photoconductive switch** [46]. (a) Optical image of an exfoliated black phosphorus layer with tapered contacts. The electrode gap is 100 nm. (b) SEM image of the photoconductive switch, showing the BP flake (black), palladium contacts (grey) and the gold coplanar waveguide (white). (c) Plot of the measured ON/OFF ratio as a function of frequency for a laser power of 50 mW and a wavelength of 1550 nm. Four different thicknesses of BP are used.

## 4. Black Phosphorus Photodetectors

Photodetection of light relies on the conversion of absorbed photons into an electrical signal. Semiconductor-based detectors (Fig.7) usually exploit a common core architecture in which the active material is electrically connected to two metallic leads (drain, D, and source, S) and the impinging radiation generates a measurable voltage or current signal. Amongst the most important detector figure of merits the responsivity (R), defined as the ratio between the magnitude of the output electrical signal (current or voltage) and the amount of incident optical power, is one commonly exploited parameter to define the sensitivity of a detector. A photoconductive detection experiment is performed applying a static *dc* bias between source and drain ($V_{DS}$) and extracting the current difference between the illuminated and dark states (Fig.7a). Conversely, a photovoltaic measurement is performed without any external bias and the output signal is measured as a voltage (current) in open (close) circuit configuration (Fig.7b).

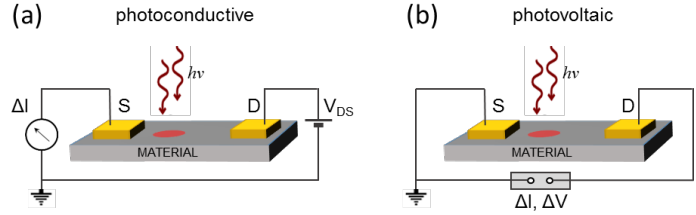

**Fig. 7: Schematic diagram of a detection experiment**. (a) Photoconductive measurement; (b) photovoltaic measurement.

The different operational regimes and detection mechanisms activated within a 2D nanomaterial mostly depend on the photon energy $h\nu$. In particular, a major distinction has to be made between the experimental condition in which $h\nu$ is larger or smaller than the energy gap ($E_g$). In the following sections, we will describe the two situations separately, first considering the implementation of BP-based detectors in the visible and telecom band ($h\nu > E_g$) and then discussing THz frequency detectors ($h\nu < E_g$).

**4.1 Photodetectors in the visible-telecom range: model**

BP is an interesting material for photo-detection stemming from the possibility to access spectral ranges otherwise prohibited to alternative 2D nanomaterials as, for example TMDs, and from strongly anisotropic optical absorption, arising from the mentioned intrinsic anisotropy in the band structure. Linear dichroism of BP has been demonstrated in the wavelength range from 400 nm to 1700 nm by performing polarization dependent absorption and reflection measurements [18] on mechanically cleaved BP flakes (Fig.8). Above the band gap, photons polarized along the armchair (x) direction are absorbed more efficiently than those polarized along the zigzag (y) direction (Fig.8a). This directly arises from the anisotropic interband transition strength: transitions between the highest valence and lowest conduction bands are allowed for the polarization along the x axis, but are partially forbidden for light polarized parallel to the y axis. This nontrivial anisotropy entails the opportunity to design conceptually new TMDs devices and applications.

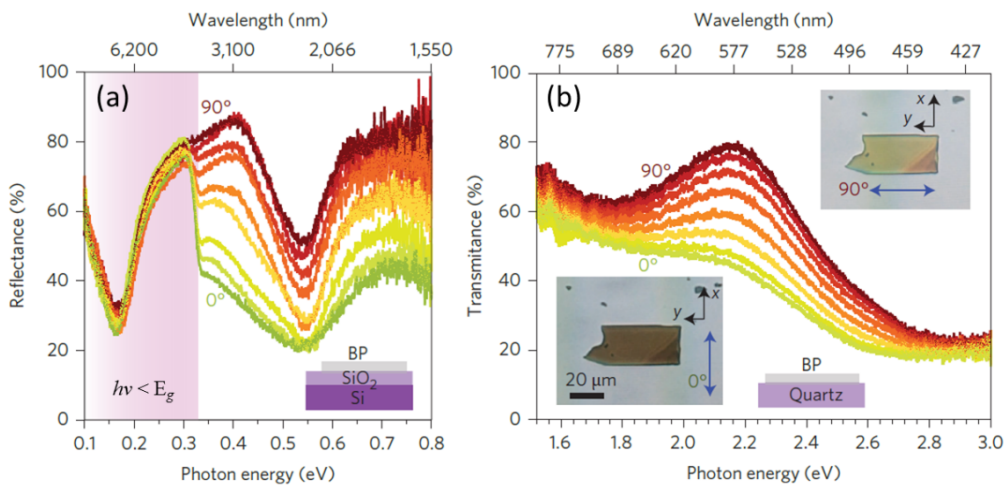

**Fig. 8: Linear dichroism in black phosphorus**. (a) Linear polarization dependence of reflectance in the infrared spectral regime, showing around 50% variation in reflectance along two perpendicular

directions for energies above the bandgap (purple shaded area). (b) Polarization dependence of the transmittance for visible light. The 0° (90°) angle corresponds to light linearly polarized parallel to the armchair (zigzag) direction. Images taken from [18].

Bulk BP has an energy gap of 0.3 eV (corresponding to ~ 4 μm). Any photon with shorter wavelength can therefore excite an electronic transition from the valence to the conduction band, thus generating an electron-hole pair. The photocurrent can therefore either be generated via electron-hole separation or driven by thermal effects. In the first case, the gained energy has to be converted into an electrical current (or voltage) before the photogenerated carriers thermalize back to the equilibrium state, without reaching the electrodes. In fact, as a consequence of losses (scattering, recombination, etc.), the number of mobile carriers decreases as they drift apart from the light spot center following an exponential decay law [18]:

$$n(r) = n_0 e^{-r/L_0} \qquad (4)$$

where $r$ is the distance from the light excitation, $L_0$ is the diffusion length, and $n_0$ is the number of electron–hole pairs generated. However, even if the photoexcited carriers does not directly reach the contacts, the stored energy can still give rise to an electrical signal mediated by a temperature increase. There are three main mechanisms through which photodetection in BP can be achieved [47]: photovoltaic, bolometric and thermoelectric effects.

The photovoltaic (PV) effect relies on the separation of photogenerated electron–hole (e–h) pairs by the built-in electric fields arising at the junctions between positively (p-type) and negatively (n-type) doped semiconductors. Upon illumination, the *internal* electric field at the junction separates the carriers and gives rise to a photocurrent at zero external bias (short-circuit current, $I_{sc}$) or a photovoltage with no current flowing (open-circuit voltage, $V_{oc}$). Since BP crystals are homogeneously doped, the typical strategy to locally induce hole and electron doping within a single flake is to employ field effect transistors (FET) endowed with multiple electrical split-gates [21] (Fig.9a), with an approach analogous to the graphene-based devices [48]. Thanks to the ambipolar nature of BP, stemming from its relatively low band-gap, this configuration allows to tune the doping to p- or n- values, depending on the applied gate voltages. Therefore, the band diagram can be controlled over the source-to-drain distance, leading to induced carrier density profiles indistinguishable from those of native p-n junctions (Fig.9b).

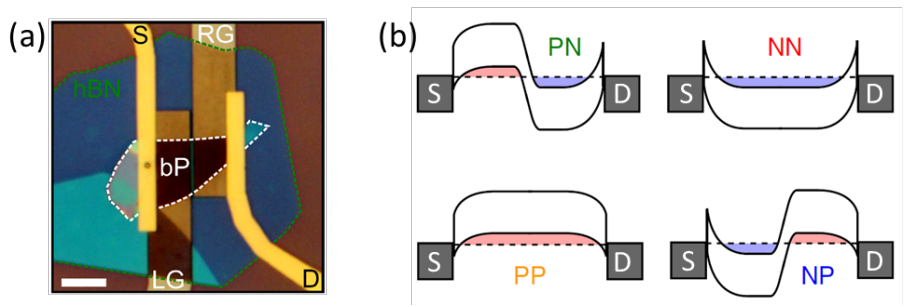

**Fig. 9: Locally gated p-n junction in black phosphorus**. (a) Optical image of a vertical back-gated structure. The left gate (LG) and right gate (RG) are electrically separated from the BP channel by a thin flake of hexagonal boron nitride (h-BN). (b) Band diagrams corresponding to the different configurations attainable by polarizing the gates. Images modified from [21].

An alternative way to separate the photogenerated electron-hole pairs is to apply a continuous *dc* bias to the contacts, to drive a photocurrent through the semiconductor [22]. This effect, known as the photoconductive (PC) effect, relies on the reduction of the electrical resistance as a consequence of the creation of extra free carriers. Since the main physics here is substantially unchanged from the p-n junction case, it can be seen as an *externally* driven photovoltaic effect. Despite its conceptual simplicity, this strategy suffers for higher levels of dark current with respect to the p-n junction configuration.

Moreover, when a voltage is applied to the active material, the measured current can be given not only by the direct collection of photoexcited carriers on the electrodes, but also by an eventual light-induced change in a macroscopic material parameter, as, for example, the mobility (μ). Indeed, simple photoconductivity measurements does not allow to distinguish between changes in the carrier mobility and changes in the number of carriers being the conductivity σ = neμ. The effect of mobility change due to uniform heating induced by photon absorption is known as the bolometric effect (BE) and its contribution is directly superimposed to the PC effect.

The origin of the bolometric effect lies in the temperature (T) dependence of the material conductance. In fact, temperature dictates both the distribution of carriers in the band structure and, for the single carrier, the probability to undergo a scattering event. The key performance indicators of a bolometer are the bolometric coefficient $\beta$ = dσ/dT, the thermal conductivity *k*, and the heat capacitance $C_h$. The thermal conductivity is the sum of the electronic and phononic contributions: k = $k_e$+ $k_{ph}$. The ratio β/k quantifies the sensitivity of the detector and the heat capacitance defines its thermal time constant.

It is worth mentioning that the bolometric coefficient has a positive sign for intrinsic, nondegenerate semiconductors, and a negative sign for extrinsic, degenerate semiconductors. Since BP is typically a degenerate semiconductor, its conductivity decreases when the temperature is raised. Then, in a photoconductive measurement (see Fig.7a), the bolometric current is negative, whereas the PC contribution is always positive. This feature have been exploited to distinguish between the two effects [3] even performing a photoconductive measurement.

In absence of a driving field, or a p-n junction, detection can still occur via the photothermoelectric effect (PTE). In 2D materials, carriers loose energy through inelastic scattering processes such as intrinsic optical and acoustic phonons, or remote surface polar phonon modes of the substrate [49]. When light is impinging on the detection area, these processes can locally heat the electronic and phononic populations, leading to re-equilibrating diffusive currents, i.e. to the Seebeck effect. The origin of this effect is found in the three main microscopic processes, in dynamic equilibrium between each other [50]. In general this mechanism can arise under two configurations: i) a temperature gradient along the conductor with a non-zero

Seebeck coefficient (S); ii) local heating at the interface between two materials with different S (thermoelectric junction or thermocouple). In both cases, a heat gradient is required in order to have a temperature gradient across one or more materials. The heat gradient can originate from either localized illumination, as with a focused laser spot with dimensions much smaller than those of the device under test, or from a strong difference in the absorption in distinct parts of the sample under uniform illumination.

In BP-based detectors the heat gradient is usually achieved by asymmetrically illuminating the surface between the electrodes [47]. In a scheme like the one reported in Fig.7, this translates in shining the radiation in proximity of one of the two contacts. Then the illuminated (*hot*) area has a larger concentration of carriers and a larger temperature than the dark (*cold*) one. If the detector is operated in an open circuit configuration a thermoelectric voltage ($V_{TE}$) can be measured:

$$V_{TE} = (S_b - S_m) \cdot \Delta T_e \tag{5}$$

where $S_b$ and $S_m$ are the Seebeck coefficients of BP and metallic leads ($S_m \approx 0$), respectively, and $\Delta T_e$ is the electron or holes temperature difference depending on the type of majority carriers.

Since BP is usually a degenerate semiconductor, $S_b$ can be parameterized in terms of the static conductivity using the Mott relation:

$$S = \frac{\pi^2 k_B^2 T}{3q} \frac{d \ln(\sigma(E))}{dE} \bigg|_{E=E_F} \tag{6}$$

where $k_B$ is the Boltzmann constant, $q = \pm e$ is the charge of the majority carriers, $\sigma(E)$ is the conductivity as a function of energy and the derivative is evaluated at the Fermi energy $E_F$.

BP has a large Seebeck coefficient at room temperature: + 335 µV/K [51]. The positive value of $S_b$ indicates a natural p-type doping of the crystal. Despite the well-known in-plane orthogonality between the dominant heat and transport directions in BP, deriving from the band structure anisotropy, $S_b$ is nearly isotropic [52]. However, the photo-thermoelectric signal is still expected to be anisotropic as a function of light polarization because of the directional dependence of the interband transition strength in the band structure of BP. This has been experimentally demonstrated in a recent work by Hong et *al.* [53] performed on 5 nm – 15 nm thick BP crystals realized by mechanical cleavage, where an increase of about 45% in the photocurrent was measured with the light polarized along the x direction.

Furthermore, the thermoelectric effect hides a more subtle anisotropy, which depends on the relative orientation between the crystalline direction and the current flow direction, parallel to the line connecting the electrodes. This can be understood from equation (5): the thermal gradient along the channel is inversely proportional the overall thermal conductivity *k*, which is larger in the y direction than in the x direction. Hence, $V_{TE}$ scales as the ratio $S_b/(k_e + k_{ph})$ [47] and the maximum detection efficiency is obtained when the BP channel is contacted along the armchair axis, where *k* is minimum.

Despite the three described mechanisms present distinctive and exclusive features, decoupling their contributions in actual experiments is a challenging task. To this end, the understanding of their coexistence and simultaneous interplay can provide a fundamental knowledge for system development and performance improvements.

A comprehensive model for uniformly doped BP photodetectors has been proposed by Low et al. [47], based on the superposition of the different effects. Taking into consideration the photoconductive system shown in Fig.7c, whenever a light is funneled on the channel, the thermal and electrical properties of the material are modified and the measured steady state current can be written as:

$$I = \sigma(V_d - V_s) + \int_0^L \sigma S_b(x) \frac{dT_e(x)}{dx} dx + \int_0^L e\mu\, n^*(x) \frac{dV(x)}{dx} dx + \int_0^L \beta[T_{ph}(x) - T_0] \frac{dV(x)}{dx} dx \quad (7)$$

The first term is the dark *dc* current ($I_{DC}$) driven by the applied voltage between the S and D contacts $V_d$ - $V_s$ = $V_{DS}$ ; the second term is the photothermoelectric current ($I_{TE}$); the third term is the photoconductive contribution ($I_{PC}$) induced by the excitation of carriers from the valence to the conduction band; the last term accounts for the bolometric contribution ($I_B$) to the current driven by the thermal excitation of the lattice ($T_{ph}$ is the phonon temperature, $T_0$ the dark-state temperature). The derivative $dV(x)/dx$ represents the static collection field along the channel.

Equation (7) does not apply to the p-n junction architecture; however, it provides a clear vision of the major differences in the light detection processes. PTE mechanism is the only one which can drive a current through the device without external bias. Its sign is uniquely determined by the temperature gradient induced by the non-uniform illumination. Conversely, bolometric effects can also arise under a homogeneous illumination (Fig.10).

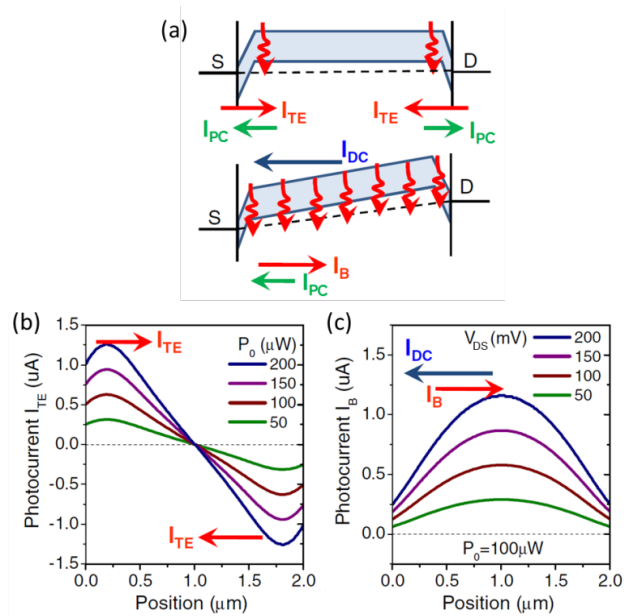

**Fig. 10: Expected photoconductive, thermoelectric and bolometric response over a 2 µm long channel**. (a) Energy band diagram of the device at zero (top) and finite (bottom) source-drain bias. The polarities of the various photocurrents, i.e., thermoelectric, bolometric, photoconductive, are indicated. (b) Simulated profile of $I_{TE}$ generated by scanning a laser across the channel at different power levels. (c) Simulated $I_B$ at different applied $V_{DS}$. Image modified from [47].

Moreover, since BP is a degenerate semiconductor, β is negative and $I_B$ has opposite sign with respect to both the dark current and $I_{PC}$. Finally, it is important to mention that, very differently from the thermoelectric case in which the thermal gradient is strongly connected with the electronic temperature, in the bolometric case, the phononic temperature is the relevant parameter which determines the scattering processes probability.

**4.2 Photodetectors in the visible-telecom range: devices**

The extensive research on BP-based devices has recently led to the development of photodetectors operating over a broadband spectral range from 400 to 1700 nm. In this section, we review present the most relevant experimental results.

The ambipolar nature of BP has been exploited to devise p-n junctions operating via local electrostatic gating (see Fig. 11) [21]. The device geometry exploits a double back-gated vertical heterostructure, in which the insulating layer consists of a thin flake of hexagonal boron nitride (h-BN). By changing the polarity of the two gates it is possible to selectively modify the channel band structure, inducing p-n, n-n, p-p or n-p alignments. This architecture allows to easily isolating the photovoltaic mechanism. Indeed, under zero-bias operation and by funneling the optical beam in the center of the channel the photoconductive, thermoelectric and bolometric effects can be ruled out.

Fig.11a,b report the gate bias dependence of the short circuit photocurrent ($I_{SC}$) and open circuit photovoltage ($V_{OC}$), measured under a 640 nm excitation wavelength. The electric output can only measured when the local gates are oppositely biased, indicating that the detection mechanism originating from the formation of a p-n or n-p junction in the channel is purely photovoltaic. The plot of the source-drain current ($I_{DS}$) as a function of $V_{DS}$ (Fig.11c) shows that $I_{sc}$ and $V_{oc}$ show a linear dependence from the incident power. The extrapolated maximum harvested power $P_{el} = I_{DS} \cdot V_{DS} \sim 13$ pW, demonstrate that the proposed BP p-n junction can be used for efficient energy harvesting.

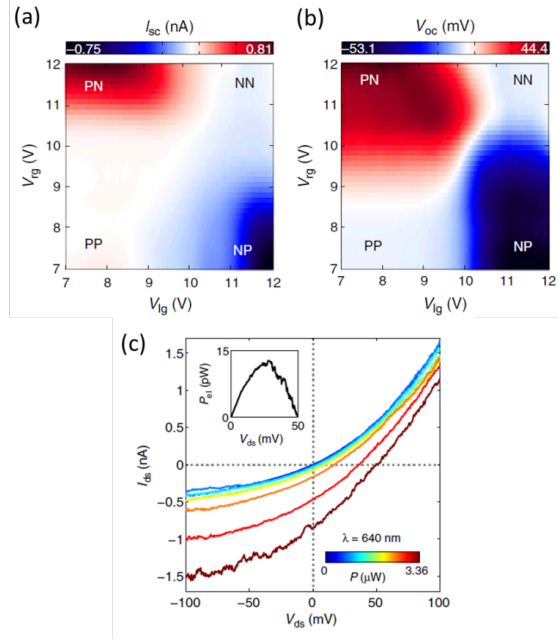

**Fig. 11:** Photovoltaic effect in a BP p-n junction for a 640 nm excitation wavelength. (a) Color map of the short circuit ($V_{DS}=0$ V) current ($I_{SC}$) as a function of the gate voltage applied to the left ($V_{lg}$) and right ($V_{rg}$) gates of Fig.9. (b) False-color map of the open-circuit ($I_{DS} = 0$ A) voltage ($V_{oc}$) as the voltages on the two local gates are changed independently. (c) $I_{DS} - V_{DS}$ characteristics of the device operated in p-n configuration under variable power. The inset shows the extracted harvested power $P_{el}$. Images taken from [18]

Photovoltaic detection effects have also been demonstrated in vertical p-n junctions [18], exploiting ionic gel gated electric-double-layer-transistors (EDLT) in which a BP flake is integrated. Under positive gate voltage, electrons can be accumulated in proximity of the BP surface, pushing holes in the bulk. The vertical junction generates the built-in electric field responsible for the PV effect and the photogenerated carriers can be collected by ring-shaped electrodes. In the same work it was found that, when no junction exists within the channel and a source-drain bias is applied, the photo-thermoelectric effect dominates the detection at zero or low *dc* voltages ($|V_{SD}| < 0.15$ V) while the photovoltaic effect starts to dominate the photocurrent generation at higher $|V_{SD}|$, in agreement with equation (7).

As a consequence of the high Seebeck coefficient of BP at room temperature, the thermoelectric effect plays a major role in the photocurrent generation and the only way to avoid its activation is to realize a completely symmetric system. This is difficult to be achieved in actual geometries since photothermoelectric current can be generated up to a micrometer away from the contacts, as a consequence of a long thermal decay length. This has been pointed out by the work of Low et *al.* [47], where the selective activation of PTE and BE in a back-gated BP phototransistor has been demonstrated. The experimental results, obtained with excitation wavelengths of 532 nm, are reported in Fig.12 and show good agreement with the theoretical model of Fig.10. The data have been collected while scanning the laser spot over the source drain distance (2.5 μm).

At zero applied bias the thermoelectric photocurrent ($I_{TE}$) changes sign when the laser is moved from left to right, as a direct consequence of the inversion of the thermal gradient. Moreover, $I_{TE}$ is not symmetric with respect to the source-drain distance, suggesting a non-uniform absorption strength within the BP flake: a strong thermoelectric signal is visible even when the center of the channel is illuminated. On the other hand, when an increasing *dc* bias is applied to the electrodes (Fig.12b) a positive bolometric term is generated along the channel, eventually overwhelming $I_{TE}$.

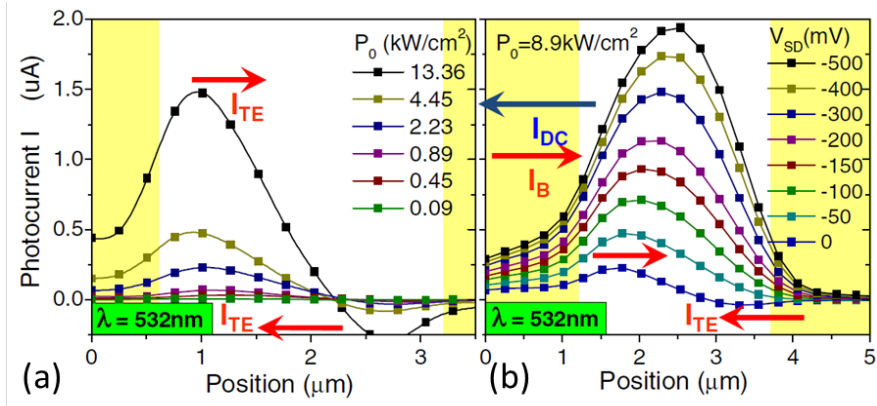

**Fig. 12:** Photocurrent profiles obtained scanning the length of a BP homogeneous channel with a 532 nm laser source. Yellow areas represent the contact areas. (a) Unbiased operation of the detector: the photocurrent is reported for different values of incident power, predominantly showing a thermoelectric behavior. (b) Photocurrent profile recorded under different biasing. The thermoelectric contribution is visible only at low biases. For $|V_{DS}| > 0.3$ V only the bolometric contribution is visible. Image modified from [47].

The present sensitivity performance of BP detectors, operating in the visible ($R_a \approx 5$ mA/W [22] and $R_a \approx 20$ mA/W [54]), are significantly lower than those of TMDs ($R_a \approx 395$ A/W in $In_2Se_3$ nanosheets [55] and $R_a \approx 10^4$ A/W in GaTe flakes [56]). In turn, BP enables faster response times ($\approx 0.2$ ms [54]) than that of TMDs, although still lower than that achieved in graphene devices (40 GHz bandwidth) [16]. However, if compared with graphene, BP detectors are less prone to high dark currents if operated in photoconductive mode.

To overcome the inherent limitations caused by intrinsic material properties (mobility, band-gap, etc.) different device architecture have been proposed. Youngblood *et al.* [3] integrate a few-layer BP photodetector into a silicon photonic circuit enhancing light-matter interaction, and reaching responsivities of ~ 130 mA/W at an incident wavelength of 1550 nm and a response bandwidth exceeding 3 GHz. The hybrid architecture is shown in Fig.11a and mimic the ones successfully exploited in graphene [57]. The BP FET was realized over a planar silicon waveguide, with a few-layer graphene as a top gate, separated from the channel by an $Al_2O_3$ gate dielectric. The integration of the active element in close contact with the waveguide allows the evanescent field to be absorbed in the material. The interaction with the excitation field mainly occurs along the in-plane direction, meaning that the absorption cross section is no more limited by the BP thickness.

The detector operates in photocurrent mode, applying a finite $V_{DS}$ and tuning the doping concentration of BP with $V_G$. Since the radiation is symmetrically fed in the detector active area, the thermoelectric response is negligible with respect to the photovoltaic and bolometric contributions, which, in turn, can be distinguished by looking at the relative sign of the photoresponse and bias voltage, as shown in Fig. 11b. No response is measured at $V_{DS} = 0$ V, as expected in absence of thermoelectric effect. At negative $V_G$, corresponding to low doping levels (electrons are the majority carriers), the current has the same sign as the applied $V_{DS}$, indicating a photovoltaic contribution. On the other hand, for positive $V_G$ (high doping), the response is switched in sign with respect to $V_{DS}$, indicating the onset of bolometric phenomena. This effect can be explained by the increasingly negative bolometric coefficient at higher carrier concentrations.

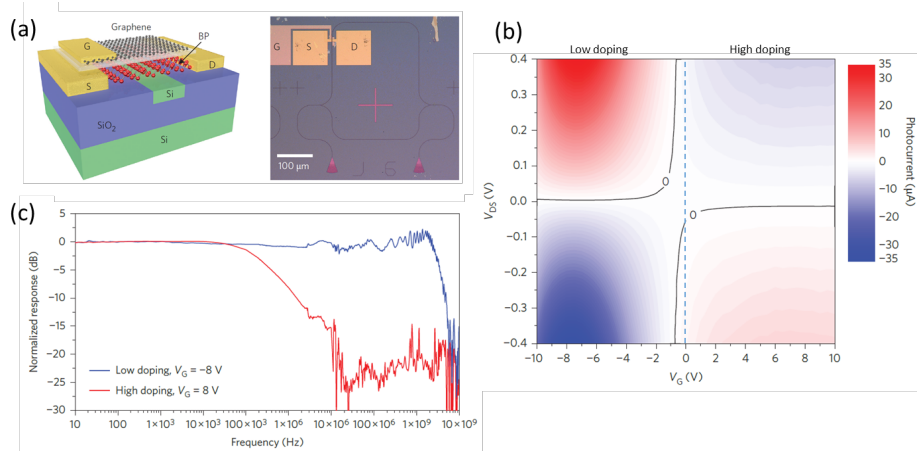

**Fig. 13: Waveguide-integrated BP photodetector**. (a) Three dimensional layout of the active element (a top-gated FET) and optical image of the complete device. The detector is integrated on a silicon photonic circuit featuring a Mach-Zender interferometer. (b) Two dimensional color map of the photocurrent as a function of the applied drain-source bias ($V_{DS}$) and gate voltages ($V_G$). Photoresponse is dominated by the photovoltaic and bolometric mechanisms in the regions of low doping and high doping, respectively. (c) Normalized photoresponse as a function of modulation frequency for low doping (blue curve) and high doping (red curve). Images taken from [3].

Interestingly, it was found that the two detection mechanisms affect not only the detector responsivity, but also its response time. Indeed, for low doping concentration the roll-off frequency is ~ 3 GHz, whereas it is only 0.2 MHz for higher doping concentrations (Fig.11c). This difference can be ascribed to the fact that the PV effect is fundamentally limited by the carrier recombination time (≈ 100 ps), whereas the bolometric time constant is given by the product of the (high [58]) in plane thermal conductivity and the heat capacitance.

Recently, BP detectors, operating in photoconductive mode, have been employed in a confocal microscope setup, for high resolution imaging has been demonstrated [54] (Fig.14). The employed target objects are 4 μm wide metallic squares deposited on glass and separated by ~2 μm. A focused laser spot is scanned over the surface of the object and the reflected signal is read by the BP phototransistor. The obtained images are shown in Figure 14c,d for wavelengths of 532 nm and 1550 nm, respectively. The large contrast allows for clear imaging with a constant contrast for sizes < 1 μm at λ = 532 nm which decreases by less than 20% when the wavelength λ = 1550 nm [54].

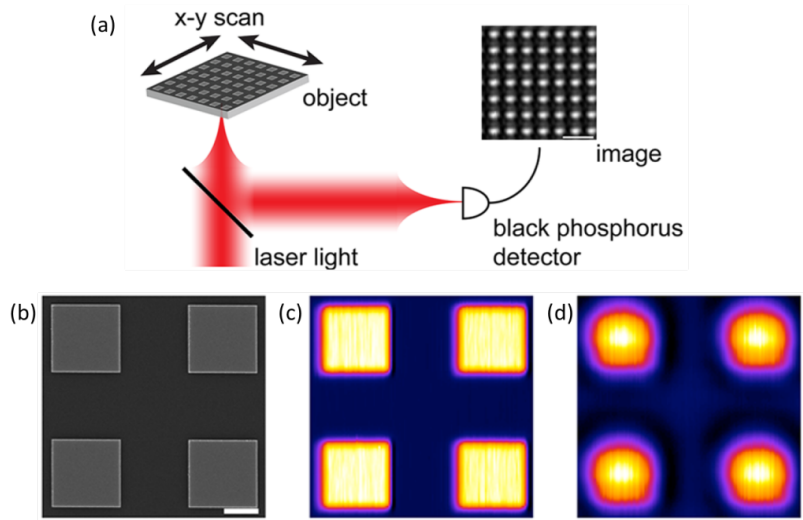

**Fig. 14:** Multispectral, high-resolution imaging with black phosphorus. (a) Schematic of the imaging process: the BP-based detector is used in a confocal microscope setup. (b) Scanning electron micrograph (SEM) of the test structure (scale bar = 2 μm). (c) Image of the test structure excited at $\lambda_{vis}$ = 532 nm. (d) Image of the test structure excited at $\lambda_{NIR}$ = 1550 nm. Image modified from [54]

Figure 15 shows a comparison between the responsivity and speed performance of a set of BP and alternative 2D-material based photodetectors (modified from ref. [59]) operating in the visible-near IR range. The performances of commercial silicon and InGaAs photodiodes are included as a benchmark. Interestingly, the device realized with BP by Youngblood et al. (ref. [3]) and the one realized with graphene by Gan et al. (ref. [57]), both employing a waveguide architecture, already present results comparable to commercially available technologies.

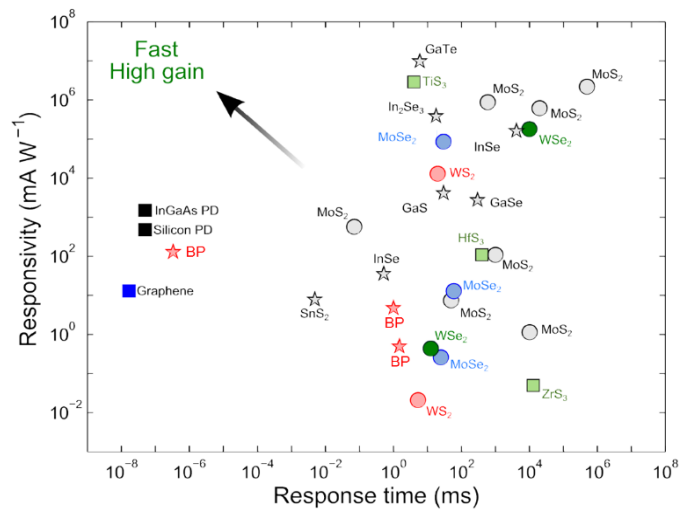

**Fig. 15:** Responsivity versus response time for most relevant performances reported in photodetectors based on 2D materials. Data about commercial silicon and InGaAs photodiodes are also reported. Image modified from ref. [59]

### 4.3 Terahertz-frequency detectors

The ability to convert light into an electrical signal with high efficiencies and controllable dynamics is a major need in photonics and optoelectronics. In the Terahertz (THz) frequency range, with its exceptional application possibilities in high data rate wireless communications, security, night-vision, biomedical or video-imaging and gas sensing, detection technologies providing efficiency and sensitivity performances that can be "engineered" from scratch, remain elusive [60]. These key priorities prompted in the last decade a major surge of interdisciplinary research, encompassing the investigation of different technologies in-between optics and microwave electronics, different physical mechanisms and a large variety of material systems [61] offering ad-hoc properties to target the expected performance and functionalities.

Commercially available direct THz detectors are based on thermal sensing elements that are either very slow (10–400 Hz modulation frequency for Golay cells or pyroelectric elements), or require deep cryogenic cooling (4 K for superconducting hot-electron bolometers), while those exploiting fast nonlinear electronics (Schottky diodes) are usually limited to low-THz frequencies for best performances [10, 62].

Moreover, stemming from the pivotal theoretical work of Dyakonov and Shur in 1996 [63] where the rectification of THz radiation in a gated two-dimensional electron gas (2DEG) was predicted, electronic devices based on the modulation induced by the incoming radiation within the conductance channel of a field effect transistor have been realized.

Photodetection of light, at THz frequencies can be accomplished by several different mechanisms like photo-thermoelectric, photovoltaic, galvanic, bolometric, plasma-wave rectification or via a combination of them [47, 64].

The detection mechanism described by Dyakonov and Shur is typically referred to as plasma-wave mechanism [65]. This effect enables the rectification of an *ac* field even at frequencies higher than the intrinsic transit time-limited cutoff of the transistor. In their present implementation, FET THz detectors conventionally operate at room temperature in a non-linear second order regime, through overdamped plasma instabilities induced along the channel when the electromagnetic *ac* field, coupled to the source and gate electrodes, simultaneously modulates the carrier density and their drift velocity [63, 66]. The resulting current exhibits a continuous *dc* component whose magnitude is proportional to the square amplitude of the *ac* field, hence to the intensity of the incoming radiation, and can be measured at the drain contact either in photocurrent mode or photovoltage mode. When charge oscillations in the channel are overdamped i.e., decay on a distance smaller than the channel length, broadband THz detection can be achieved.

Conversely, a strong resonant photoresponse is expected in materials having plasma damping rates lower than both the frequency ν of the incoming radiation and the inverse of the electron transit time in the channel. To fulfill the above condition, mobilities of at least several thousand cm$^2$/Vs and operating frequencies >1 THz need to be exploited, so that the transistor can operate in the so-called high frequency regime ($2\pi\nu\tau > 1$), where $\tau$ is the momentum relaxation time.

Room temperature FET THz detectors are typically realized in high electron mobility transistors (HEMT) [67] or CMOS technology [68], leading to cost-effective applications and opening the possibility of implementing multipixel focal plane arrays. More recently, this approach has been extended to one-dimensional (1D) and 2D nanostructures, like semiconductor InAs and InN NWs [66, 69, 70], graphene [71, 72] and phosphorene [60, 73], obtaining efficient detection from 300 GHz to 3 THz.

In all these implementations, a huge effort has been devoted to efficiently couple the free-space electromagnetic wave ($\lambda$ = 100 µm - 1 mm in the 300 GHz – 3 THz range) to the deeply sub-wavelength active element of the FET. A typical approach is the integration of the transistor in a planar antenna, with the aim of funneling the electromagnetic energy on the rectifier. The antenna can be exploited to provide the system with the necessary asymmetry in order to create a preferential direction for the current flow, once the plasma waves are excited within the channel. Indeed, the plasma wave mechanism is predicted to be more efficient when the radiation is coupled between the S and G electrodes (Fig.16). This allows FET detectors to work without any applied bias, thus reducing the noise level and the system complexity.

Despite the fact that the photon energy is not sufficiently large to excite optical transitions and then to generate extra free carriers within the channel, the electromagnetic energy can still be converted into thermal energy by the currents driven on the antenna surface. Hence, thermal effects can still occur in the material, giving rise to thermoelectric and bolometric contributions to the photocurrent. This issue has been recently addressed engineering the architecture of antenna-coupled THz nanodetectors exploiting thin flakes of exfoliated BP, to selectively activate each of those processes, individually. The inherent electrical and thermal in-plane anisotropy of BP was exploited to selectively control the detection dynamics in the BP channel, at room-temperature and with state-of-the art detection efficiency [60],

The BP THz detectors demonstrated so far exploit the integration of a $SiO_2$-encapsulated single BP flake in an antenna-coupled top-gate (G) FET. $SiO_2$ is here of crucial importance to avoid oxide reactions and material degradation due to the environmental instability.

In their first implementations, BP FETs have been integrated in planar dipole bow-tie antennas, which present a broadband response with limited impedance. Single-crystalline ingots of BP were grown via a chemical vapor transport technique similar to the one reported in Ref. [74]. Flakes having thickness in the range 8-14 nm were then mechanically exfoliated from bulk BP crystal using a standard adhesive tape technique on a 300 nm thick $SiO_2$ layer on the top of a 300 µm-thick intrinsic silicon wafer. The exfoliated flakes were initially identified via optical and scanning electron microscopy (SEM) and then characterized with atomic force microscopy (AFM), and linearly polarized micro-Raman spectroscopy, to determine the layer thickness and the crystallographic directions through the intensity ratio of the $A_g^2$ and $A_g^1$ active modes, respectively. Thin flakes with thicknesses $h \sim$ 9-14 nm were then individually contacted with proper adhesion layer/metal sequences to define the S and D FET electrodes via aligned electron beam lithography (EBL) (see Methods). Three sets of samples have been devised, all showing identical architectures but a different crystallographic orientation of the BP flake along the FET channel axis. In samples A (FET channel along the x-axis) and C (FET channel along the D-axis) the S and G electrodes were then patterned in the shape of a half 110° bow-tie antenna, introducing a strong asymmetry in the mechanism of light harvesting; conversely, in sample B (FET channel along the y-axis), the same 110° bow-tie antenna was symmetrically placed at the S and D electrodes (see Fig. 16)

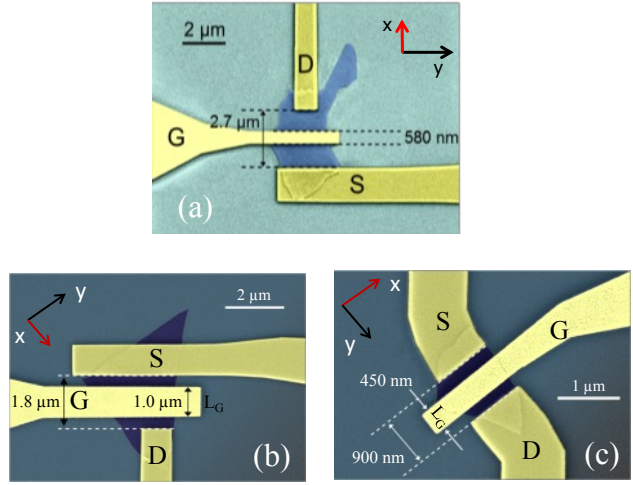

**Fig. 16:** Scanning Electron Micrographs of THz black phosphorus photodetectors. (a) Flake contacted along the *x* direction. (b) Flake contacted along the 45° direction. (c) Flake contacted along the *y* direction.

Figures 17a,d,g shows the responsivity ($R_v$)/photovoltage ($\Delta u$) trends plotted as a function of $V_G$ in the three devised architectures measured, via a lock-in acquisition technique when a 0.29 THz beam was impinging on the devices at $V_{SD} = 0$ V.

In the case of fully asymmetric FET architectures (sample A and B), two main effects are expected to be triggered by the incoming THz radiation: (*i*) the excitation of plasma oscillations along the FET channel [63, 71]; (*ii*) the heating of the metallic contacts due to the THz-driven local currents inside the antenna arms [72]. By contrast, symmetric geometries, like the one employed for sample C, can likely induce bolometric detection effects, under specific material/geometry configurations. Both plasma-wave and thermoelectric effects are conversely prevented by the inherent device symmetry.

The plasma-wave rectification effect, triggered by the antenna asymmetric radiation feeding in the conductive channel, results in an asymmetric charge density modulation, which will in turn induce a longitudinal electric field, with a preferential direction for the current flow. Under this regime (diffusive overdamped plasma-wave self-mixing regime), the generated photoresponse can be deduced from the transfer characteristics of the FET via the relation [63, 71, 75]:

$$\Delta u_T \propto -\frac{1}{\sigma} \cdot \frac{d\sigma}{dV_G} \cdot \left[\frac{R_L}{\frac{1}{\sigma}+R_L}\right] \quad (8)$$

where $R_L$ is the finite impedance of the measurement setup including the readout circuitry.

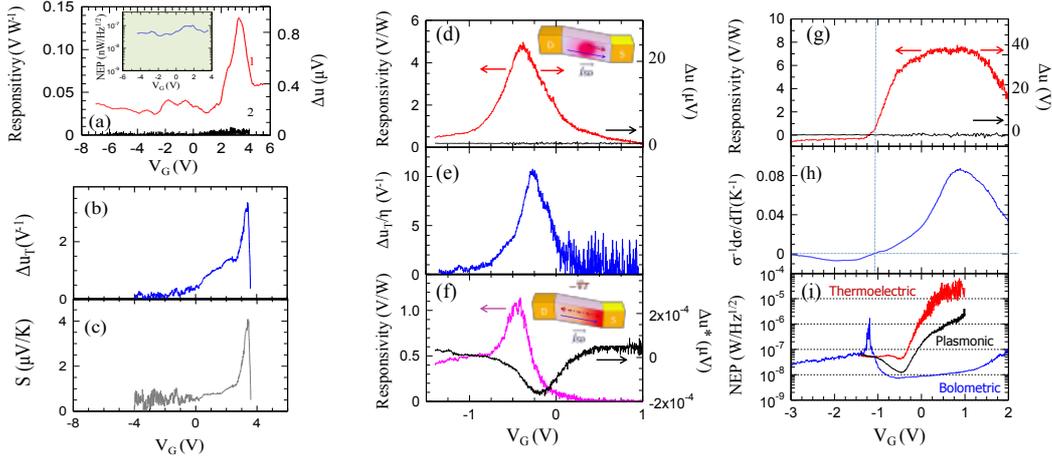

**Fig. 17: Terahertz detection**. (a,b,c) Performance of the detector exploiting BP flakes oriented in the x direction (sample A). (a) Gate bias dependence of the experimental room temperature responsivity ($R_v$)/photovoltage ($\Delta u$). The red line (curve 1) was measured by impinging the THz beam on the detector surface (THz-on); the black line (curve 2) was measured while blanking the beam with an absorber (THz-off). Inset: Noise equivalent power as a function of $V_G$, extracted from the relation $N_{th}/R_v$. The noise spectral density ($N_{th}$) has been calculated via the relation $N_{th} = (4\, k_B T \sigma^{-1})^{1/2}$. (b) Predicted photoresponse as a function of $V_G$, under the overdamped plasma-wave regime. (c) Seebeck coefficient extracted from the measured transfer characteristics (Mott relation). (d,e,f) Performance of the detector oriented in the 45° direction (sample B). (d) Gate bias dependence of the experimental room temperature responsivity ($R_v$)/photovoltage ($\Delta u$). The red line was measured with THz-on, the black line was measured with THz-off. Inset: schematics of the overdamped plasma-wave dynamics. (e) Predicted photoresponse as a function of $V_G$, under the overdamped plasma-wave regime; (f) responsivity and photovoltage extracted from photocurrent measurements, while detuning the THz frequency from the antenna resonance; (g) Gate bias dependence of the experimental room temperature responsivity ($R_v$)/photovoltage ($\Delta u$) in sample c; (h) Predicted bolometric trend; (Noise equivalent power (NEP) as a function of $V_G$, for the plasma-wave (sample B, 0.29 THz), thermoelectric (sample B, 0.32 THz) and bolometric (sample C) detectors.

Equation (8), results in the predicted photovoltage trends shown in Fig. 17b and 17e. The comparison with the corresponding experimental $\Delta u$ curves (Fig. 17a and 17d, right vertical axis), shows good agreement: $R_v$ peaks at positive and negative $V_G$, respectively, without any sign switch, as expected, and in full agreement with the predicted behavior. However, in the case of sample A, $R_v$ does not decrease to zero (Figure 17a), in clear contrast with the model predictions, but saturates at an average value of ~ 0.03 V/W, suggesting contributions of thermoelectric origin, arising from the presence of the ungated p-doped BP regions, and subsequent formation of p-p-p junctions, as already shown in graphene THz detectors [71]. Figure 17c, shows the Seeback coefficient trend, extrapolated from Equation 6, which identifies photo-thermoelectric effect as the dominant photo-detection process. Such a conclusion is well supported by the fact that being the material thermal conductance anisotropic and since the current is flowing along the armchair axis of low thermal conductance, our geometry significantly enhances BP thermoelectric performance [52]. The comparison between $\Delta u$ and the photovoltage value extracted from on/off photocurrent measurements $\Delta u^*$ [61] usually provides a valuable way to unveil the dominant effect that is contributing to the detection. Under the assumption that thermoelectric effects dominate, the THz-induced carrier distribution gradient generates a diffusive flux of holes from the *hot*-side (S) to the *cold*-side (D) of the channel, hence, under zero-bias operation, (positive) charges will be accumulated at D whose potential will rise from zero to a (positive) thermoelectric value $\Delta u_{pe}$. Conversely, if $V_{SD} \neq 0$ (with $V_D > V_S$ being S grounded) a certain amount of current will flow through the channel. If the sample is kept in the dark, the only electromotive force will be provided by the *dc* voltage $V_{SD}$, and $I_{SD,off}$ (Fig. 17f, inset) will flow from D to S, i.e. in the opposite direction with respect to the light induced photothermoelectric current ($I_{pe}$) generated. On the other hand, if over-damped plasma-wave effects dominate in our device B, the excited carrier density would be pushed toward one channel side or the other depending on the structure asymmetry. In fact, at zero bias ($V_{SD} = 0V$), the sign of the plasma-wave photovoltage $\Delta u_{pw}$ is not known a priori. However, the corresponding $\Delta u^*_{pw}$ is well defined [61], since the applied *dc* voltage $V_{DS}$ sets an asymmetric direction across the channel that will cause the charge density to drift towards the S side. The THz-induced current will then sum up with the pre-existent *dc* current, leading to $\Delta u^*_{pw} > 0$, in agreement with our experimental data [61]. These considerations support the conclusion that sample A behaves like a plasma-wave THz detector, operating in the non-resonant overdamped regime [71]. However, the detuning of the impinging frequency provided to be a valuable solution to activate detection dynamics of purely thermoelectric origin [61]

To elucidate the nature of the detection process in sample C, we estimated the bolometric photovoltage ($V_B$), via its functional dependence [47] from the ratio $\frac{1}{\sigma}\frac{d\sigma}{dT}$:

$$V_B = \frac{I_B}{\sigma} \propto \frac{\gamma}{\sigma} = \frac{1}{\sigma}\frac{d\sigma}{dT} \qquad (3)$$

The trend reported in Fig. 4(d) is in excellent agreement with the experimental responsivity curve, thus confirming that device B behaves like a bolometer, as expected if one consider that the longitudinal in-plane acoustic phonons show a sound speed (then a conductance) along the *y*-direction (8397 m/s) almost twice than the sound speed along the *x*-direction (4246 m/s).

Remarkably, maximum $R_v$ of 5.0 V/W (in a plasma wave architectures) and 7.8 V/W (in a bolometric architecture) have been reached, significantly larger than those reported in exfoliated graphene FETs [76], leading to noise equivalent power levels down to 7 nW/√Hz, 10 nW/√Hz and 45 nW/√Hz for the BP-bolometer, plasma-wave and thermoelectric detector, respectively [61, 73].

More recently, by reassembling the thin isolated atomic planes of hexagonal borum nitride (hBN) with a few layer phosphorene (black phosphorus (BP)) we mechanically stacked hBN/BP/hBN heterostructures to devise high-efficiency THz photodetectors operating in the 0.3-0.65 THz range from 4K to 300K with a record SNR = 20000.

hBN, a III-V compound with an energy band gap of ≈ 5.2-5.4 eV constitutes an ideal encapsulating material for BP. Being impermeable to gases and liquids it provides permanent shielding to ambient exposure, leading to extremely air-stable devices [77]. Furthermore, it enables reaching record mobility (up to ≈ 1350 cm$^2$V$^{-1}$s$^{-1}$ at room temperature [78] and ≈ 4000 cm$^2$V$^{-1}$s$^{-1}$ at 1.5 K) thanks to its flatness and its compatibility with honeycomb structure.

In contrast with oxide-encapsulated structures, [79] when stacked over other layered crystals, hBN indeed forms clean and inert interfaces, preventing charge traps or dangling bonds. Under this configuration, hBN can also be seen as a valuable *ready-to-use* gate dielectric, due to its high breakdown voltage ($\approx 1$ V nm$^{-1}$, depending on the number of layers and on the carrier type) [80] and the sufficiently large dielectric constant (3.5-4), [81,82] which in turn allows high gate-to-channel capacitance value required for tuning the device transport and optical properties. Moreover, it is in principle capable to provide a large mobility increase at low temperatures ($\approx$ an order of magnitude) with respect to any oxide-surrounded device ($\approx$ a factor of 2) [83].

## Acknowledgements

The authors acknowledge support from the European Union Seventh Framework Programme grant agreement n° 604391 Graphene Flagship, the European Union through the MPNS COST Action"MP1204 TERA-MIR Radiation: Materials, Generation, Detection and Applications".